\documentclass{ws-ijmpb}

\usepackage{graphicx,amssymb,amsmath}

\begin{document}

\markboth{F. Baldovin and E. Orlandini}
{Nonequilibrium behavior of long-range Hamiltonians}

%
\catchline{}{}{}{}{}
%

\title{NOS\'E-HOOVER AND LANGEVIN THERMOSTATS DO NOT REPRODUCE  
  THE NONEQUILIBRIUM BEHAVIOR OF LONG-RANGE HAMILTONIANS}


\author{FULVIO BALDOVIN and ENZO ORLANDINI}

\address{
Dipartimento di Fisica and
Sezione INFN, Universit\`a di Padova,\\
Via Marzolo 8, I-35131 Padova, Italy\\
baldovin@pd.infn.it, orlandini@pd.infn.it}



\maketitle

\begin{history}
\received{Day Month Year}
\revised{Day Month Year}
\end{history}

\begin{abstract}
We compare simulations performed using the Nos\'e-Hoover and the
Langevin thermostats with the Hamiltonian dynamics of a long-range
interacting system in contact with a reservoir. 
We find that while the statistical mechanics equilibrium properties of
the system are  recovered by all the different methods, the
Nos\'e-Hoover and the  Langevin thermostats fail in reproducing the
nonequilibrium behavior of 
such Hamiltonian. 
\end{abstract}


\section{Introduction}
Long-range characterizes the interactions of a number of different
physical systems like, e.g., plasmas, wave-matter 
systems, gravitational
systems, Bose-Einstein condensates.\cite{dauxois} 
In these cases, the customary assumptions of
statistical mechanics are put into question because of the 
inapplicability of the Boltzmann transport equation.\cite{balescu}
In fact, nonequivalence between the microcanonical and the
canonical ensemble approaches,\cite{touchette} nonergodicity and
topological nonconnetctivity\cite{borgonovi} has been detected for
long-range system.
Classical long-range Hamiltonian models assume then a central role
in order to compare the dynamical behavior of macroscopic 
phase functions like the system's energy, its temperature, or its
magnetization, with the correspondent
predictions of statistical mechanics. 
It is known that the dynamics of long-range Hamiltonians
displays long-living quasi-stationary states (QSS) in 
microcanonical ($\mu$C) simulations, i.e., when the system is
isolated.   
This aspect has been studied in details by several groups in the last
decade.\cite{qss_literature}
On the other hand, at least in terrestrial-scale experiments,
the system cannot be considered isolated.  
It is then interesting to see if QSSs are reproduced in more
``experimental'' settings, especially in view of some theoretical
results,  based on the Langevin equation, that seem to rule out such a
possibility.\cite{choi}

In Ref.~\refcite{hmf_can,hmf_qss} we addressed this issue by introducing a
Hamiltonian setup in which the 
long-range system is coupled
with a thermal reservoir through microscopic interactions. 
We discovered the persistence of long-lasting QSSs whose life-time
depend on the system size and on the coupling strength between the system and
the reservoir.\cite{hmf_can,hmf_qss} 
In this Paper we further investigate this point by comparing standard
methods for the simulation of  
a thermal bath interacting with the system, 
namely the Nos\'e-Hoover (NH) and the Langevin (LA) thermostats,
with the above Hamiltonian.

An important observation with respect to the NH and the
LA schemes is that both algorithms implicitly assume some equilibrium
features.  
In the first case, a single degree of freedom with ``effective mass''
$Q$ is added to the system in order to simulate a thermal bath. 
This additional
degree of freedom has the capability of adsorbing and releasing an
arbitrary amount of energy, and its temporal
scale is appropriately redefined in such a way to generate the
Boltzmann-Gibbs equilibrium canonical distribution for the
system.\cite{frenkel} 
The parameter $Q$ can be used for tuning a better convergence of the
algorithm. 
At difference, the Langevin approach is based on the assumption of a
well defined separation between the time-scales of the dynamics of the
diffusive particle (slow dynamics) and that of the underlying thermal bath
(fast dynamics). Because of this, the bath is assumed to
be in thermal equilibrium at all the integration steps and via the
equipartition theorem the damping and the stochastic coupling constant
characterizing the diffusive behavior
are related by a specific, temperature-dependent, 
fluctuation-dissipation relation.\cite{frenkel}
As a consequence of these implicit assumptions, it is not guaranteed
that the NH and the LA integration strategies can be safely applied
out of equilibrium, especially for a system 
which is known to produce unconventional effects
in $\mu$C simulations.

In the following, we take an ``empirical'' attitude by
implementing the NH and the LA integration schemes for the simulation
of a long-range Hamiltonian system in contact with a thermal bath.
We find that while the equilibrium behavior of the system is
equivalently recovered by the different approaches, the NH and the LA
thermostats  do not properly account for the nonequilibrium features
of the Hamiltonian dynamics.
 
\begin{figure}[bt]
\centerline{
\includegraphics[width=0.70\columnwidth]{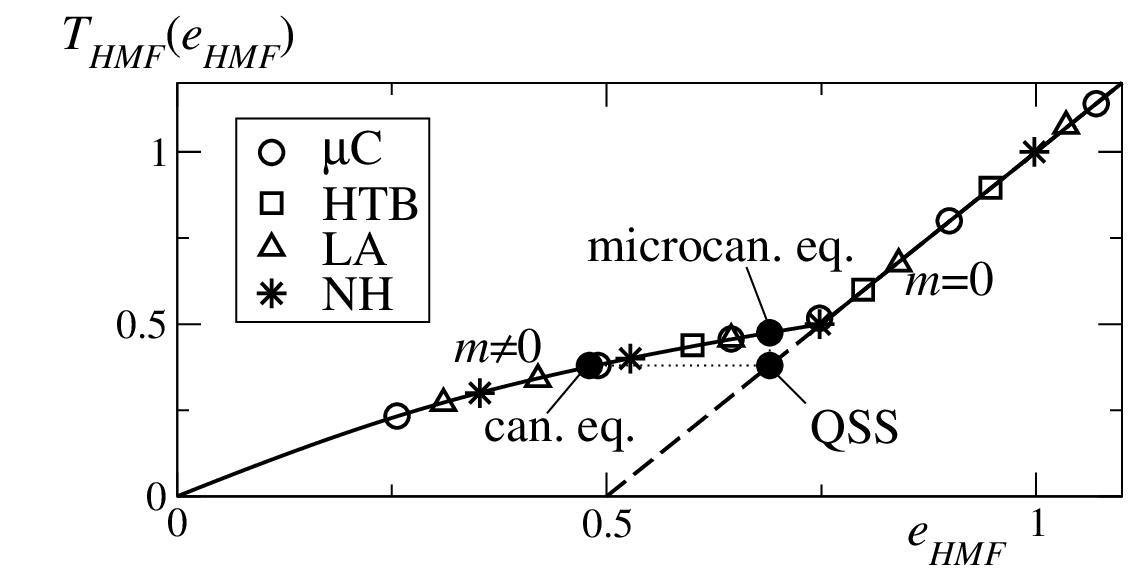}
}
\vspace*{8pt}
\caption{
  Caloric curve. The solid line is the Boltzmann-Gibbs
  equilibrium solution and the dashed line is the prolongation of the
  ordered  phase to subcritical energies.
  Empty symbols are the average value of
  $e_{HMF}(t)$ at equilibrium.
  Full circles correspond to the QSS studied in the paper
  and to the subsequent microcanonical and canonical equilibrium 
  obtained as $t\to\infty$.
}
\label{caloric}
\end{figure}

\section{Equilibrium simulations}
The long-range interacting Hamiltonian considered in
Refs.~\refcite{qss_literature,choi,hmf_can,hmf_qss} is called
Hamiltonian Mean Field (HMF) model and it can be thought as a  
set of $M$  globally coupled $X Y$-spins 
with Hamiltonian
\begin{equation}
H_{HMF}=K_{HMF}+V_{HMF}=\sum_{i=1}^M\frac{l_i^2}{2}
+\frac{1}{2M}\sum_{i,j=1}^M\left[1-\cos(\theta_i-\theta_j)\right],
\label{hmf}
\end{equation}
where $\theta_i\in[0,2\pi)$ are the spin angles 
assumed with unit momentum of inertia
and $l_i\in\mathbb R$ their angular momenta (velocities). 
Its equilibrium statistical mechanics solution predicts
an high-energy disordered phase separated from a low-energy ordered one by
a second order transition occurring at the specific energy 
$e_{HMF}\equiv E_{HMF}/M=0.69$ (we use dimensionless units).
The order parameter is the magnetization of the system $m_{HMF}\equiv
|\sum_{i=1}^M(\cos\theta_i,\sin\theta_i)|/M$, and the presence of the
kinetic term endows the spin system with a proper Hamiltonian dynamics
in which one can define the temperature $T$
as twice the specific kinetic energy,
$T_{HMF}\equiv2K_{HMF}/M$. 
Notice the relation $e_{HMF}=(T_{HMF}+1-m_{HMF}^2)/2$.

The Hamiltonian thermal bath (HTB) considered in
Refs.~\refcite{hmf_can,hmf_qss} is given by the full Hamiltonian system
$H=H_{HMF}+H_{TB}+H_I$, 
\begin{eqnarray}
H_{TB}&=&\sum_{i=M+1}^{N}\frac{l_i^2}{2}
+\sum_{i=M+1}^{N}\left[1-\cos(\theta_{i+1}-\theta_i)\right],\\
H_{I}&=&\epsilon\sum_{i=1}^{M}\sum_{s=1}^S\left[1-\cos(\theta_{i}-\theta_{r_s(i)})\right], 
\end{eqnarray}
where $H_{TB}$ and $H_I$ are respectively the Hamiltonian of the
thermal bath and that of the interaction between HMF model and thermal
bath ($N\gg M$). 
As the coupling constant $\epsilon$ vanishes, the $\mu$C
dynamics of the HMF is recovered (see Ref.~\refcite{hmf_can}
for details). 

A similar circumstance is valid for the LA thermostat where
the equation of motion for the HMF model are
\begin{equation}
\ddot{\theta_i}=-\gamma\dot{\theta_i}-\sum_{j=1}^M\sin(\theta_i-\theta_j)+
\sqrt{2\gamma T}\;\xi_i(t),\quad
i=1,2\ldots,M.
\label{langevin}
\end{equation}
In Eq. (\ref{langevin}) $\xi_i(t)$ is a Gaussian white noise
characterized by a zero average $\langle\xi_i(t)\rangle=0$ and
correlation 
$\langle\xi_i(t)\xi_j(t^\prime)\rangle=\delta_{i j}\delta(t-t^\prime)$. 
Indeed, in the limit $\gamma\to0$, Eqs. (\ref{langevin}) reduce to
the $\mu$C Hamiltonian equations of the HMF model. 

In Fig. \ref{caloric} we present the results of the simulations in the
different setups at equilibrium. 
These simulations are obtained by setting initial conditions
close to equilibrium for the HMF model. Specifically, we used a Maxwellian
distribution of velocities and an initial value of $m_{HMF}$ and
of $T_{HMF}$ close to those of equilibrium for the given fixed energy
($\mu$C) or thermal bath temperature (HTB, NH and LA).
After the relaxation to equilibrium,
we verified that all the different dynamics give the same results 
for the values of the phase functions
$e_{HMF}$, $T_{HMF}$ and $m_{HMF}$.

\begin{figure}[bt]
\centerline{
\includegraphics[width=0.45\columnwidth]{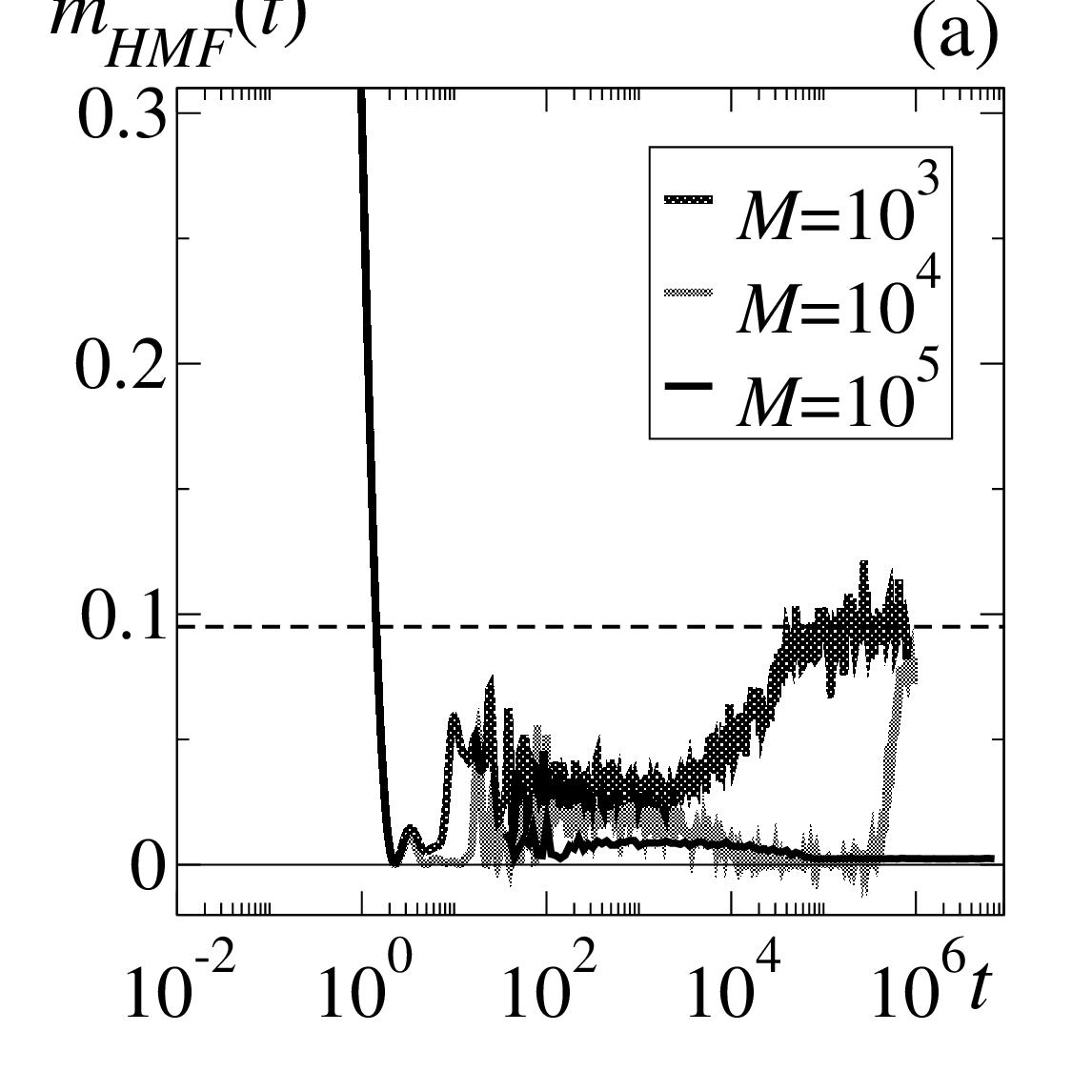}
\includegraphics[width=0.45\columnwidth]{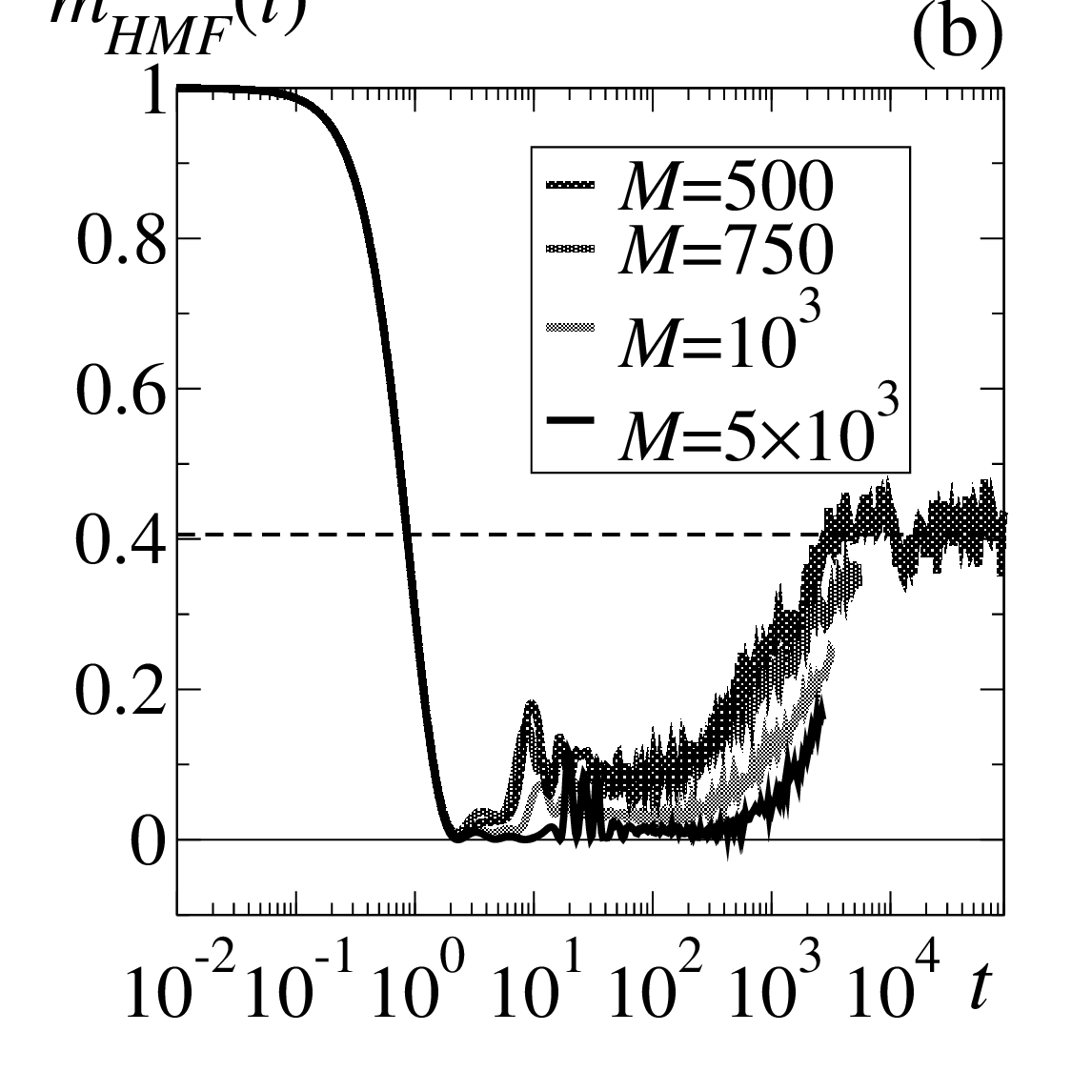}
}
\vspace*{8pt}
\caption{
  QSSs in $\mu$C simulations (a) and in HTB simulations with
  $\epsilon=0.01$ (b). Dashed lines are the equilibrium values for
  $m_{HMF}^2$. 
  See Refs.~7,8 for further details.  
}
\label{mag_can}
\end{figure}

\section{Nonequilibrium simulations}
The nonequilibrium analysis is performed by changing the initial
conditions for the dynamical variables of the HMF model, while keeping
all the other parameters unchanged with respect to the equilibrium
results. 
Consistently with other studies reporting the
existence of QSSs,\cite{qss_literature}
we have chosen at time $t=0$ 
a delta
distribution for the angles [$p_{HMF}(\theta,0)=\delta(0)$ so
that $m_{HMF}^2(0)=1$] and a uniform distribution for the
velocities [$p_{HMF}(l,0)=1/2\bar l,\;l\in[-\bar l,\bar l]$,
with $\bar l\simeq2.03$].
In this way, the initial energy of the HMF model is set to the
subcritical value $e_{HMF}(0)\simeq0.69$.

The $\mu$C simulations reveal in this case the existence of a 
{\it violent relaxation process\cite{qss_literature}} (for a time of order
$1$) followed by a QSS which can be displayed e.g. by plotting 
the time dependence of $m_{HMF}^2$ (Fig. \ref{mag_can}a).  
The QSS life-time
diverges in the thermodynamic limit $M\to\infty$ and in this limit
$m_{HMF}^2$ vanishes.  
The same kind of results are obtained using the HTB, although now the
QSS life-time diminishes as $\epsilon$ increases
(Fig. \ref{mag_can}b). 
Notice that in the $\mu$C simulations the system relaxes to
equilibrium at fixed energy, whereas in the HTB ones the relaxation is
at fixed thermal bath temperature. 
This produces a consistent difference in the
equilibrium values of $m_{HMF}^2$ (see also
Fig. \ref{caloric}).

\begin{figure}[bt]
\centerline{
\includegraphics[width=0.45\columnwidth]{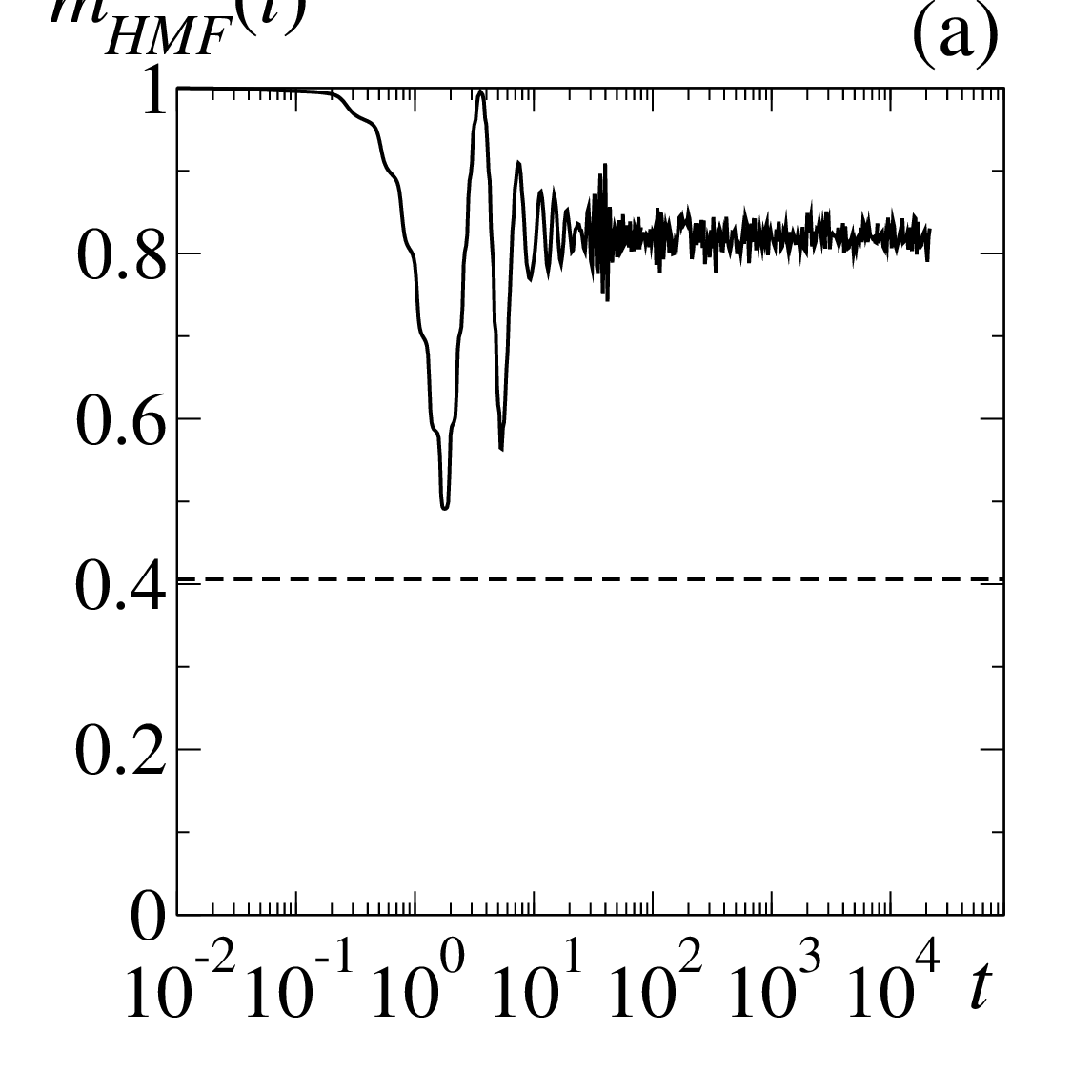}
\includegraphics[width=0.45\columnwidth]{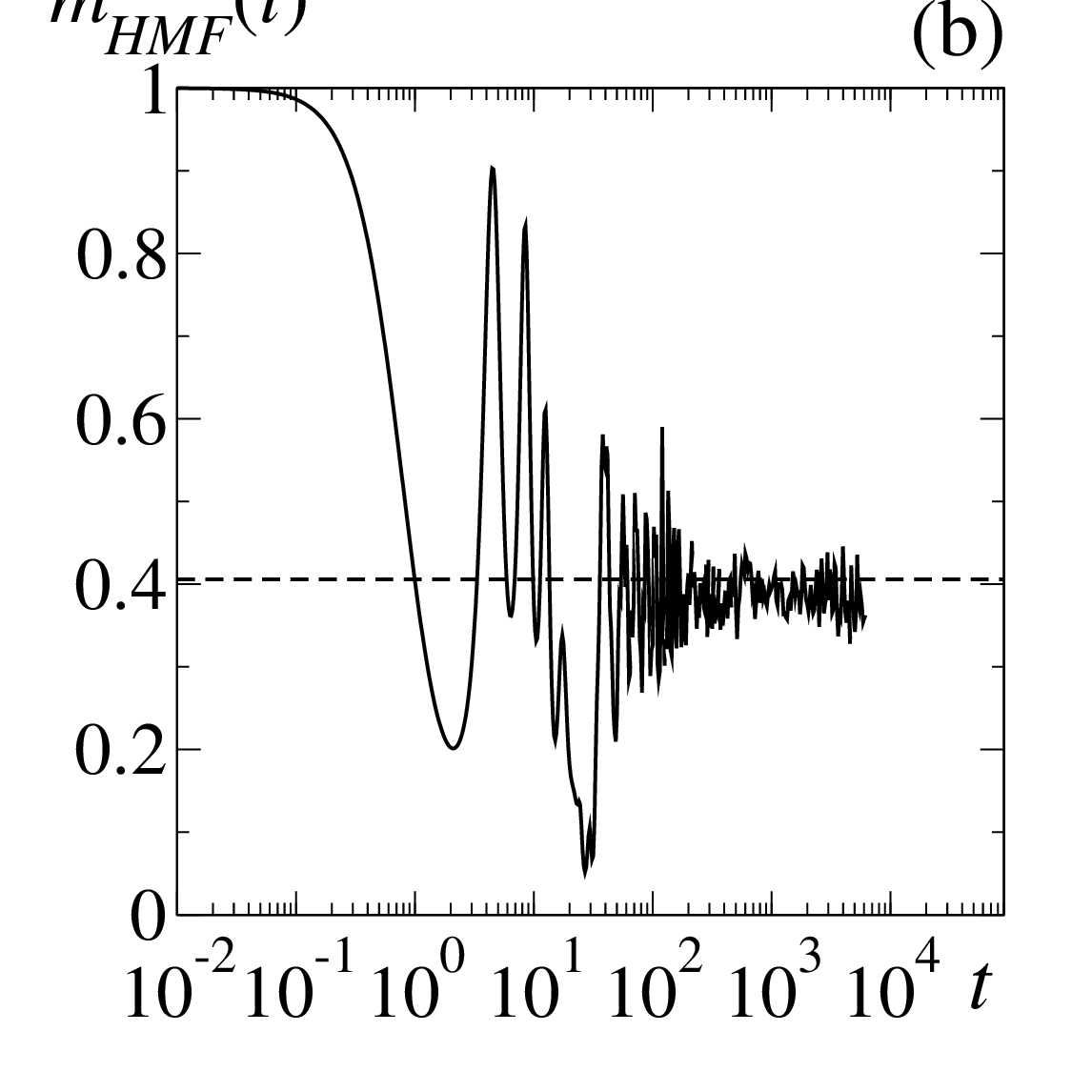}
}
\vspace*{8pt}
\caption{
  NH simulations with far-from-equilibrium initial conditions. 
  (a) With $Q=1$ the magnetization does not converge to equilibrium. 
  (b) With $Q=M=10^3$ the magnetization converges to equilibrium
  without displaying QSSs. 
}
\label{mag_nh}
\end{figure}

Unlike the results in the previous section, a NH
integration scheme implemented with nonequilibrium initial
conditions for the HMF model does not always guarantee the final
convergence to equilibrium (Fig. \ref{mag_nh}). 
We found that only when $Q$ is larger then the value
of the system size $M$ the convergence to equilibrium is
realized. Still, the nonequilibrium dynamics is characterized by
fluctuations of the phase functions which display no relation with the 
Hamiltonian simulations. Another drawback of the NH method is that if
initially the HMF model has vanishing total momentum, 
this quantity remains zero during all the integrations
steps. 

\begin{figure}[bt]
\centerline{
\includegraphics[width=0.45\columnwidth]{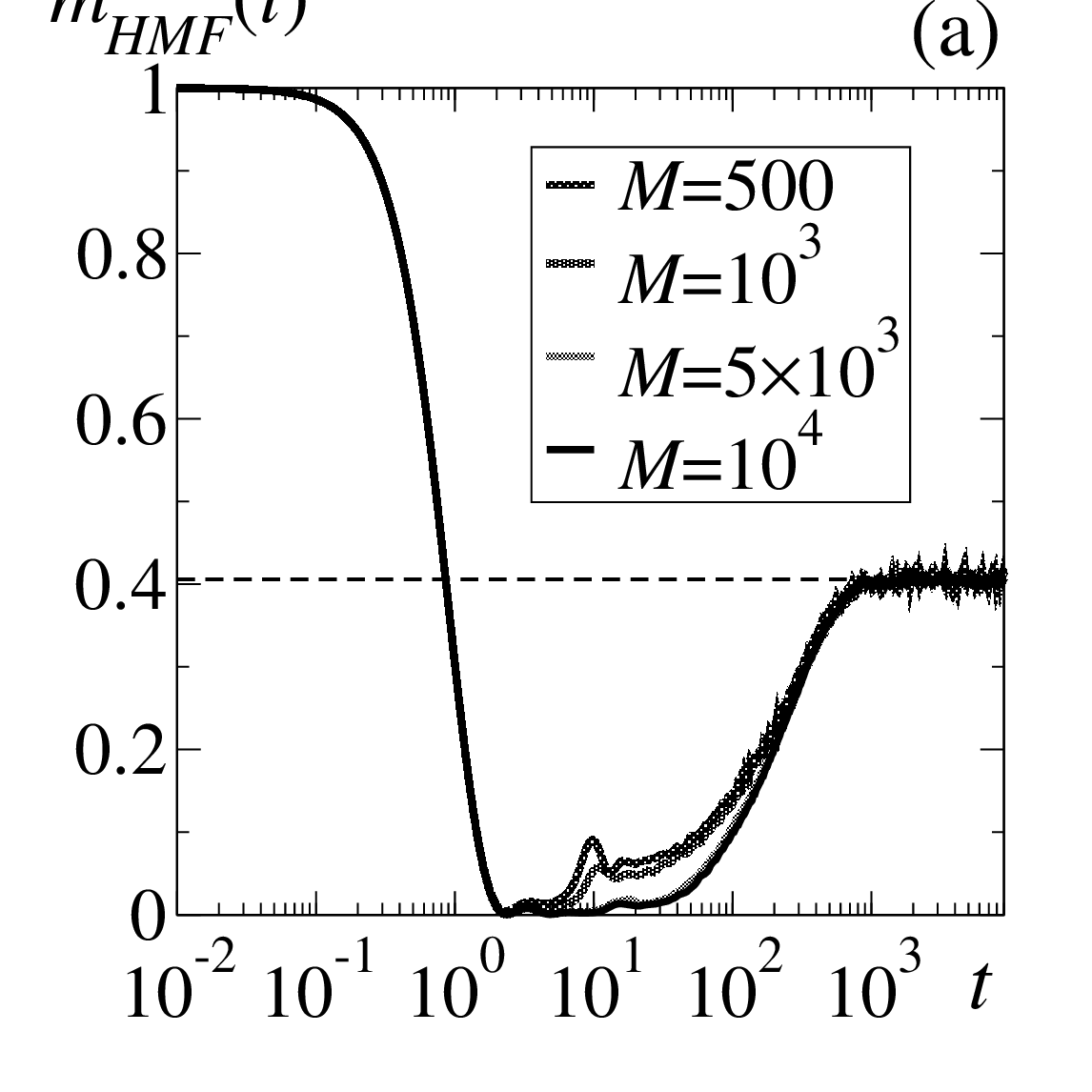}
\includegraphics[width=0.45\columnwidth]{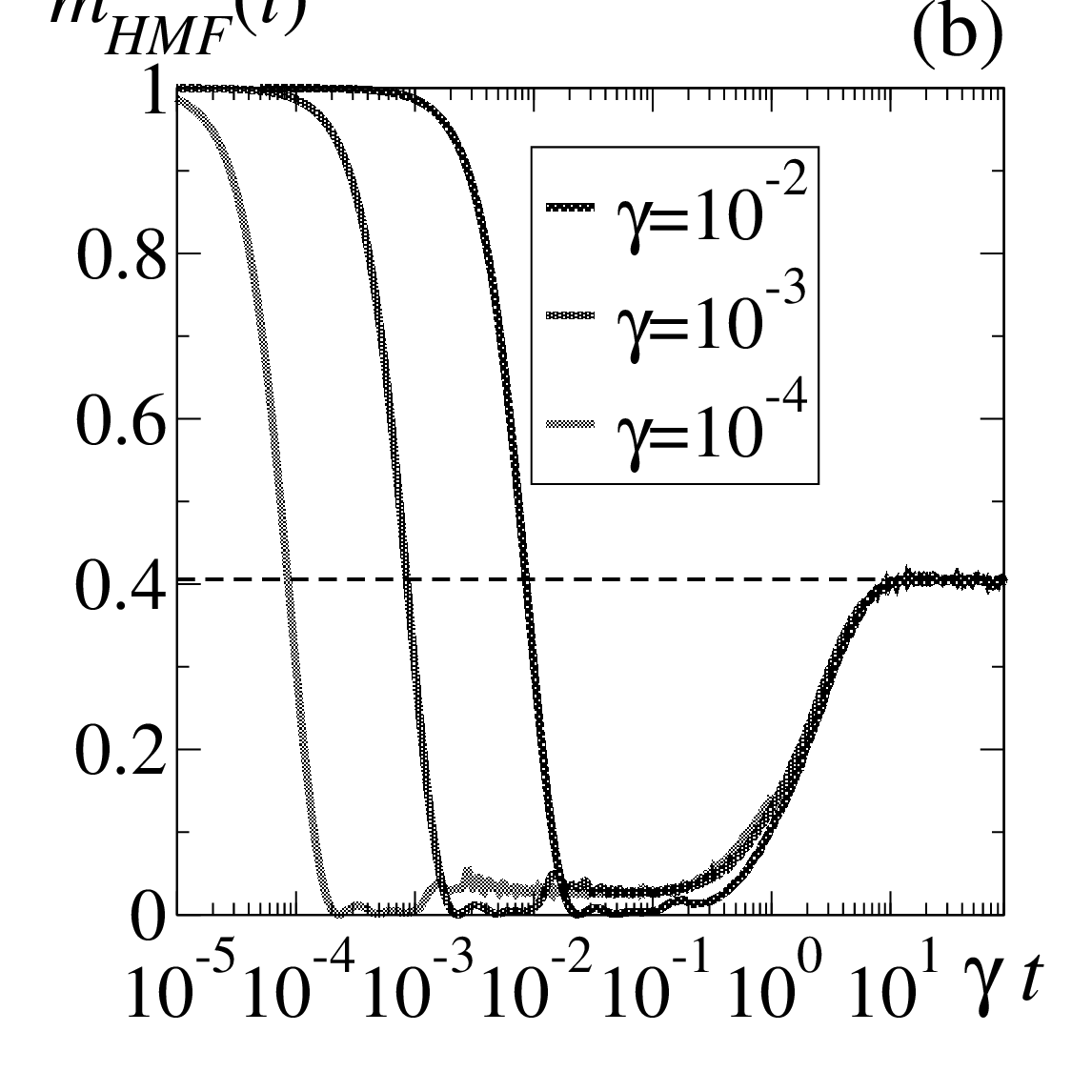}
}
\vspace*{8pt}
\caption{
  LA simulations with far-from-equilibrium initial conditions. 
  (a) As $M$ increases, the QSS's life-time remains constant.
  (b) The crossover time between QSS and equilibrium scales as 
  $\gamma t$.
}
\label{mag_lan}
\end{figure}

The analysis of the LA simulations reveals some interesting new
results. In this case, as for the HTB, the convergence to equilibrium
is observed for any value of $\gamma>0$ and  the total momentum of
the HMF model fluctuates during the simulation, as it is expected.
Also, the violent relaxation process is coherently reproduced by the
LA simulations and a QSS follows for which $m_{HMF}^2\to0$ as $M$
grows. 
Nonetheless, the QSS life-time appears to be independent from the
system size $M$ for any value of $\gamma>0$ (Fig. \ref{mag_lan}a).  
This life-time also shows an interesting dependence on
$\gamma$. While the violent-relaxation time is of order $1$
independently on $\gamma$, the crossover time from the QSS to the
equilibrium scales as $\gamma t$ (Fig. \ref{mag_lan}b).
This scaling law implies an infinite
life-time of the QSS in the $\mu$C limit $\gamma\to0$, independently
on $M$. 
Since such a result is in contrast with purely Hamiltonian $\mu$C
simulations (Fig. \ref{mag_can}a), it suggests the presence of a
discontinuity in $\gamma=0$.   

It is interesting to recall that a stability analysis of
the Fokker-Planck equation derived from Eq. (\ref{langevin}) shows
that anomalous, non-Maxwellian, velocity probability density functions 
are (neutrally) stable only in the $\mu$C limit
$\gamma=0$.\cite{choi}
The somehow unexpected\cite{choi} 
presence of QSSs in LA simulations may be
related to the fact that during the QSS the HMF model
does  not thermalize with the thermal bath (see
Ref.~\refcite{hmf_qss,preparation} for details). 

In conclusion, by showing a specific example in which the NH and the
LA thermostats simulations do not agree with the correspondent fully
Hamiltonian ones, 
our findings constitute a general warning against the straightforward
application of equilibrium-based algorithms for the description of the
statistical nonequilibrium behavior. 

{\bf Acknowledgments.} FB acknowledges the organizers of the 
``International Conference on the Frontiers of Nonlinear and Complex
Systems'', Hong Kong, May 2006, for generous support.

\end{document}